\documentclass{article}
\usepackage{ijcai23}

\usepackage{times}
\usepackage{soul}
\usepackage{url}
\usepackage[hidelinks]{hyperref}
\usepackage[utf8]{inputenc}
\usepackage[small]{caption}
\usepackage{graphicx}
\usepackage{amsmath}
\usepackage{amsthm}
\usepackage{booktabs}
\usepackage{algorithm}
\usepackage{algorithmic}
\usepackage[switch]{lineno}

\nolinenumbers

\title{Do Orcas Have Semantic Language? \\[4mm]
Machine Learning to Predict Orca Behaviors Using Partially Labeled Vocalization Data}

\author{Sophia Sandholm
\affiliations
Shady Side Academy
\emails
ssandholm14@outlook.com
}

\begin{document}

\maketitle

\begin{abstract}
\textit{Orcinus orca} (killer whales) exhibit complex calls. They last about a second. In a call, an orca typically uses multiple frequencies simultaneously, varies the frequencies, and varies their volumes. Behavior data is hard to obtain because orcas live under water and travel quickly. Sound data is relatively easy to capture. As a science goal, we would like to know whether orca vocalizations constitute a semantic language. We do this by studying whether machine learning can predict behavior from vocalizations. Such prediction would also help scientific research and safety applications because one would like to predict behavior while only having to capture sound. A significant challenge in this process is lack of labeled data. We work with recent recordings of McMurdo Sound orcas~\cite{Wellard20:Cold} where each recording is labeled with the behaviors observed during the recording. This yields a dataset where sound \textit{segments}---continuous vocalizations that can be thought of as call \textit{sequences} or more general structures---within the recordings are labeled with superfluous behaviors. Despite that, with a careful combination of recent machine learning techniques, we achieve 96.4\% classification accuracy. This suggests that orcas do use a semantic language. It is also promising for research and applications.
\end{abstract}

\section{Introduction}

%
Marine biologists have recordings of \textit{Orcinus orca} (killer whale) vocalizations. They have identified in the recordings what they coin ``\textit{calls}'' that last around a second each. Even within an individual call, an orca typically uses multiple frequencies simultaneously, varies the frequencies, and varies their volumes. 

To our knowledge, it is not known whether calls or longer sound \textit{segments}---call \textit{sequences} or even more general structures---constitute a \textit{semantic} language. That is, do orcas communicate meaning? 

Researchers have clustered orca calls into 12, 28, or 78 call types depending on the study~\cite{Schröter19:Segmentation,Wellard20:Cold,Ford_1984}. However, there is little understanding of what, if anything, these calls convey~\cite{Schröter19:Segmentation}. This is because orcas live mostly under water and move quickly, so the sound recordings are rarely accompanied by other data that would support such reasoning. We use a recent orca sound recording collection that is rare in the sense that it has such auxiliary data~\cite{Wellard20:Dataset,Wellard20:Cold}. In particular, it has partially labeled behavior data. By carefully combining and tailoring select modern machine learning techniques, we show that the sound segments can be used to predict orca behavior with 96.4\% accuracy. This strongly suggests that orcas use a semantic language. 

To our knowledge, this paper is the first to use partially labeled learning to study animal vocalizations, first to use machine learning to analyze orca sound segments beyond individual calls, and first to predict orca 
behavior from their vocalizations. Prior research on whale sounds has primarily focused on identifying whales in passive acoustic listening and identifying individual call types. For example, Bergler \textit{et al.}~\shortcite{Bergler19:Deep} used unsupervised learning to cluster orca calls. Bergler \textit{et al.}~\shortcite{Bergler19:Deep1} worked on classifying orca calls using a ResNet-18 neural network. Bergler \textit{et al.} ~\shortcite{Bergler19:ORCA-SPOT} created a system using convolutional neural networks that can differentiate orca calls from environmental noise. Beyond orcas, studies of those kinds on whale vocalizations using machine learning have been conducted on false killer whales~\cite{Murray98:The}, sperm whales~\cite{Jiang18:Clicks,Bermant19:Deep,Andreas22:Toward}, long-finned pilot whales~\cite{Jiang18:Clicks}, right whales~\cite{Shiu20:Deep}, beluga whales~\cite{Zhong20:Beluga}, fin whales~\cite{Best22:Temporal}, humpback whales~\cite{Allen21:A}, and blue whales~\cite{Miller22:Deep}. Also, the PAMGUARD software has been developed to identify cetacean presence in passive acoustic listening data~\cite{Gillespie09:PAMGUARD}. 

The recent sound recordings of Wellard \textit{et al.}~\shortcite{Wellard20:Dataset,Wellard20:Cold} that we use last between 51 seconds and 41 minutes each, for a total of 3.5 hours of recordings of orcas from the
McMurdo Sound in Antarctica. Wellard \textit{et al.} labeled each recording with all the orca behaviors that they observed during that recording: T for traveling, F for foraging, S for socializing, and M for milling/resting. Each recording can therefore have a \textit{combination} of behavior labels. 

We \textit{segment} these recordings to isolate continuous orca vocalizations that are typically many times as long as individual calls. In a segment, the orcas may not be exhibiting all of the behaviors that were observed during the recording from which the segment was taken. This yields a data set where the segments are labeled with superfluous behaviors. This presents a significant problem: a lack of fully labeled behavior data. We use a custom loss function which is designed for learning on partially labeled data to combat this issue. That, combined with our tailored sound preprocessing pipeline and the use of a slightly modified ResNet-34 convolutional neural network trained on that custom loss function, enables us to achieve 96.4\% classification accuracy on the test sets, that is, on sound segments that our system has not seen during training. 

Beyond suggesting that orcas use a semantic language, creating a deep learning system that can identify behavior just by vocalizations can be a powerful tool for marine biologists studying marine ecology and animal behavior. In addition, it can be helpful to scientists studying whale language. It can also have safety applications and other uses. It may also help get us closer to mankind's desire to be able to communicate with animals.





The rest of the paper is organized as follows. 
In Section~\ref{se:sound_proc_pipeline} we describe our sound preprocessing pipeline. As data, we use recent orca recordings by Wellard \textit{et al}. We segmented these recordings and transformed the recordings into resampled decibel Mel spectrograms. We used them as inputs to our machine learning system. In Section~\ref{se:deep_learning}, we describe our deep learning system. We use a slightly modified pre-trained ResNet-34 convolutional neural network and a custom loss function to achieve our classification task. In Section~\ref{se:results}, we present our experimental results. In Section~\ref{se:examples_of_what_the_system_learned}, we demonstrate what our algorithm learned by showing examples of the classification.  We present conclusions and possible avenues for future research in Section~\ref{se:conclusions}.

\section{Sound Preprocessing Pipeline}
\label{se:sound_proc_pipeline}
We used recent recordings of the Type C Ross Sea orcas by Wellard \textit{et al.}~\shortcite{Wellard20:Cold}. They collected the recordings on the McMurdo Sound in Antarctica using a hydrophone. They recorded the orcas nine times throughout December, 2012 and January, 2013. There were two days, January 8 and January 11, where the orcas were recorded at two separate locations on the same day. With each recording, they also  documented the behaviors that the orcas exhibited during the recording. After analyzing the recordings, they also documented which call types were used by the orcas in each recording. 


We accessed the recordings through the Dryad Digital Repository~\cite{Wellard20:Dataset}. Based on the four possible behavior labels in the data---\textit{traveling (T)}, \textit{foraging (F)}, \textit{socialising (S)}, and \textit{milling/resting (M)}---the recordings could in principle have any one of the $2^4 -1 = 15$ possible label combinations. However, in reality each recording had one of the following six label combinations: \{T\}, \{F, S\}, \{T, S\}, \{T, F\}, \{T, F, M\}, or \{T, F, S, M\}. The raw data files in the database are not labeled directly with their behavior labels. Instead, they are only labeled with dates. We extrapolated the data labels associated with each recording by comparing the behaviors observed on each recording day and the date labels on the sound files~\cite{Wellard20:Dataset}. For January 8th and January 11th, where the orcas were recorded at two separate locations on the same day and a different set of behaviors was observed during each recording, we used the individual calls (documented by Wellard \textit{et al.}~\shortcite{Wellard20:Cold}) observed in each file to identify which recording was associated with a given label combination. To do this, we used an audio software called Audacity to look at the spectrogram of the sound and identify the presence of certain call types which only occurred during a given behavior label combination.

At the heart of our study is the analysis of sound \textit{segments}---call \textit{sequences} or even more general structures---not just individual calls. For this purpose, we \textit{segmented} the recordings to create the sound segments for analysis by our system. We conducted the segmentation manually via Audacity using the following rules to define a segment.
\begin{itemize}
\item Each segment had to be longer than half a second. 
\item Each segment had to occur at least two seconds apart from other orca vocalizations. If vocalizations occurred less than two seconds apart, we considered them part of the same segment.
\item The orca vocalizations in any segment needed to be seen on the spectrogram in Audacity and be audible when played back. 
\end{itemize}
The numbers of sound segments after segmentation, with their associated label combinations, are shown in Table~\ref{ta:frequencies}. 
\begin{table}[!h]
    \centering
    \begin{tabular}{cc}
        \hline
        Label combination  & Number of sound segments \\
        \hline
        \{T\} & 124 \\
        \{F, S\} & 122 \\
        \{T, S\} & 7 \\
        \{T, F\} & 95 \\
        \{T, F, M\} & 112 \\
        \{T, F, S, M\} & 58 \\
        \hline
        Total & 518 \\
    \end{tabular}
    \caption{Numbers of sound segments with the various label combinations.}
    \label{ta:frequencies}
\end{table}

\begin{figure*}[!t]
    \title{Sound Processing Pipeline}
    \includegraphics[height =8.1cm, width=\textwidth]{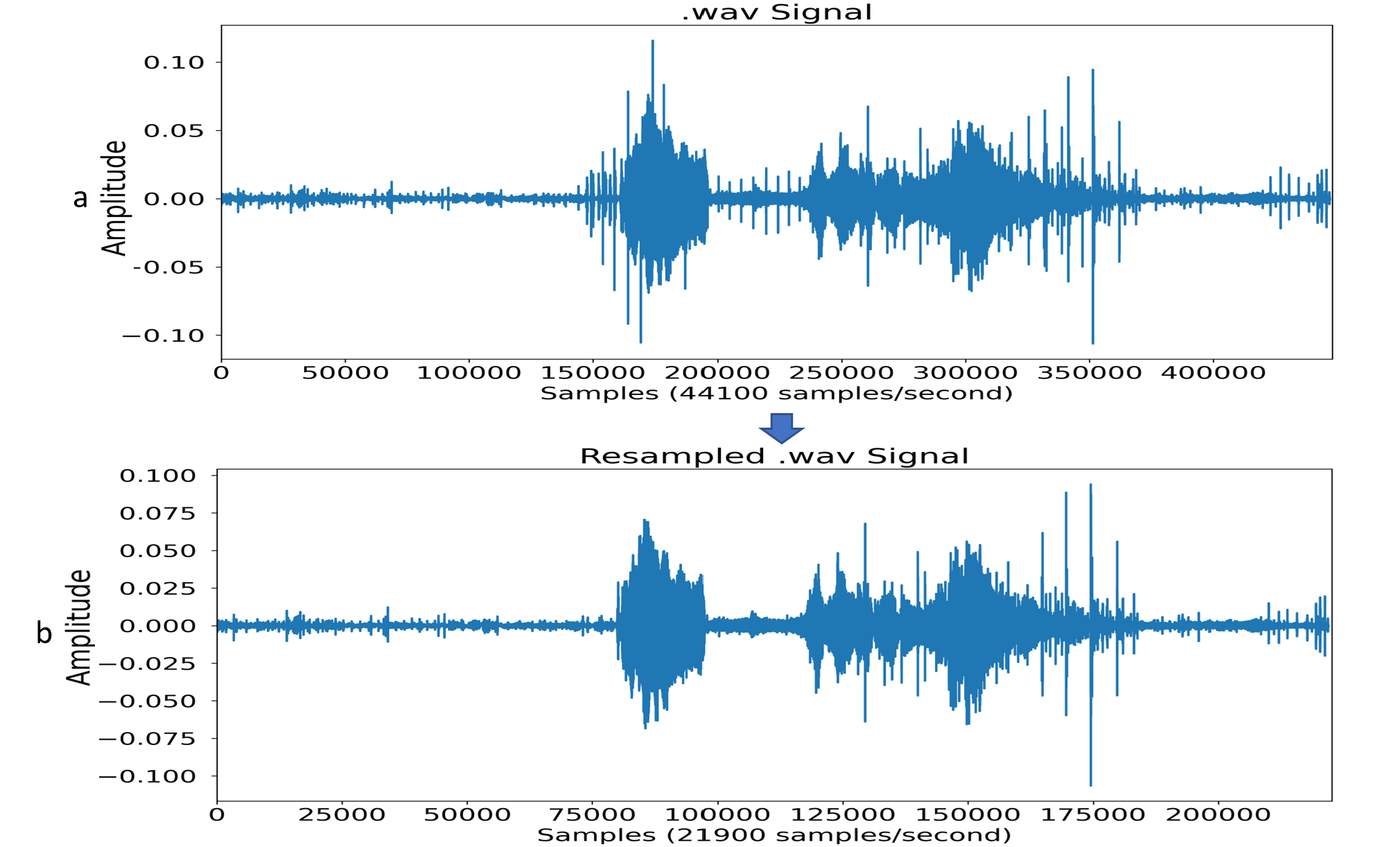} 
    \includegraphics[height =10.7cm, width=\textwidth]{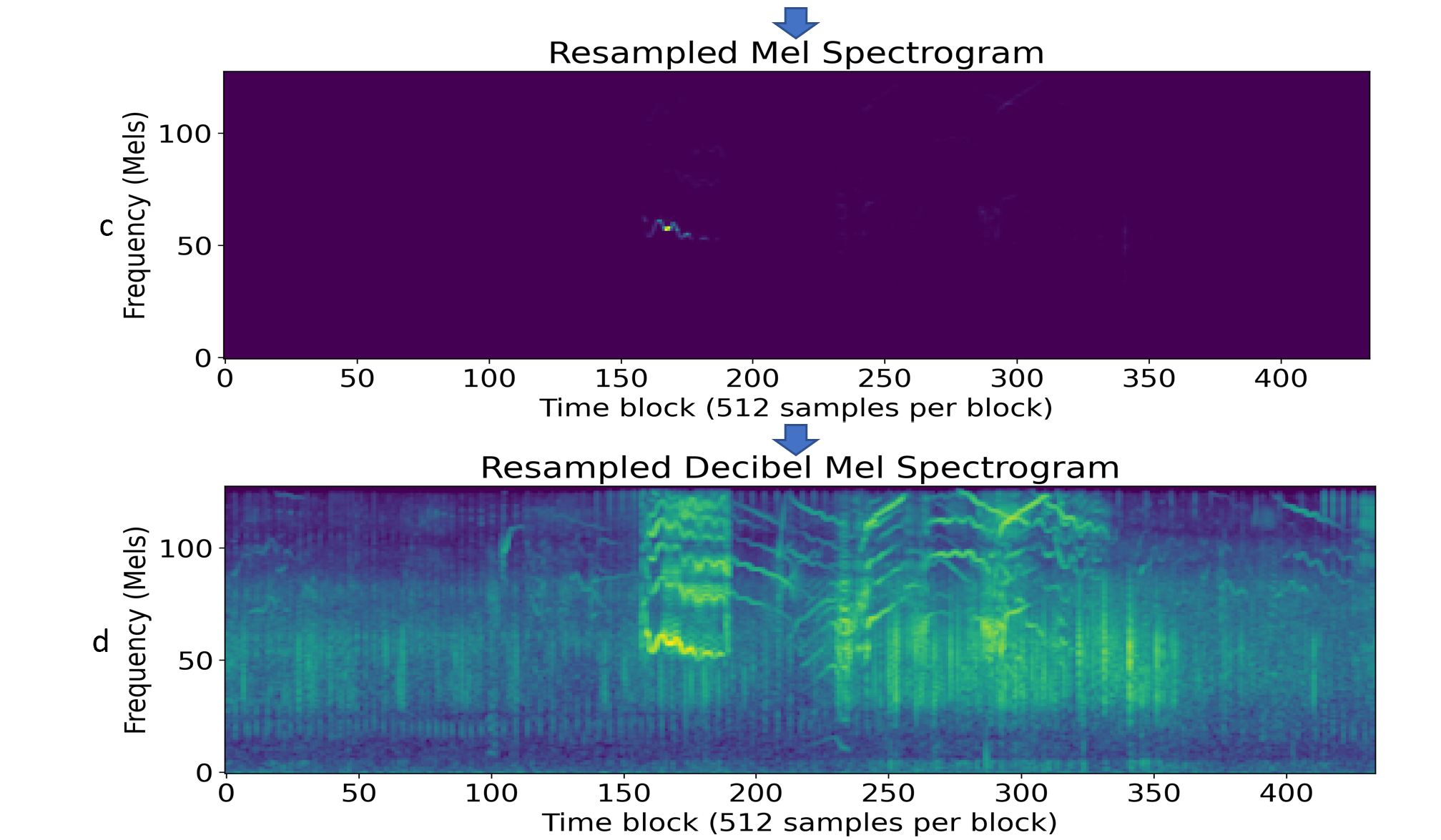} 
    \caption{Key steps of the programmatic part of our sound processing pipeline illustrated on a 10-second orca sound segment. \textbf{a)} Raw recorded sound signal (that is, .wav format) that shows pressure (amplitude) at the sampling points. \textbf{b)} Resampled .wav signal. \textbf{c)} Resampled Mel spectrogram. \textbf{d)} Resampled Decibel Mel spectrogram. (The padding is not shown in these figures. It would show as a dark uninteresting solid on both sides of the spectrogram. The normalization of the decibel Mel spectrogram to make a decibel Mel spectrogram image is also not shown. The decibel Mel spectrogram image looks the same as the decibel Mel spectrogram.)}
    \label{fi:soundpros}
\end{figure*}

Next, we wrote a program to resample the segments so that every segment ended up with a sampling rate of 21,900 samples per second. See Figure~\ref{fi:soundpros} (a) and (b). 
Then, we padded the segments (with constants) before the beginning and after the end so that all of the segments are the same length. We padded the segments rather than truncating them all to the length of the shortest segment so that no data in any segment was lost.

Then, in order for our deep learning system to effectively analyze the segments, we transformed the waveforms of the segments into \textit{decibel Mel spectrogram images} as follows. We first transformed the waveforms into Mel spectrograms. A spectrogram takes the Fast Fourier Transform at every time block (\textit{i.e.}, time window); we used a time block of 512 samples/time block. It ends up being a picture with time blocks on the x-axis, frequency in Hertz (Hz) on the y-axis, and strength of each particular frequency as the intensity of the (monochrome) color.  This enables one to not only study frequencies that are present in the sound but changes in the frequencies across time. Then, to get a Mel spectrogram, one transforms the frequency axis nonlinearly so that human hearing perceives the same change in tone for each additive change in Mels. See Figure~\ref{fi:soundpros} (c). Then, we transformed the Mel spectrograms to decibel Mel spectrograms by putting the color intensity (\textit{i.e.}, strength of each frequency) on a log scale.
As shown in Figure~\ref{fi:soundpros} (d), this transformation makes the patterns dramatically more noticeable. Finally, we normalize the decibel Mel spectrograms to have values between 0 and 255, which creates a normalize decibel Mel spectrogram image. It actually does not look any different than the decibel Mel spectrogram. We used these images as inputs for our deep learning system.

\section{Deep Learning System and Dealing with Superfluous Behavior Labels}
\label{se:deep_learning}
Our deep learning system leverages a pre-trained ResNet-34 convolutional neural network and a custom loss function. ResNet-34 is a modern convolutional neural network architecture that includes ReLU units and skip connections~\cite{He16:Deep}. It has 34 layers which alternate between convolutional layers and pooling layers. The final output layer is a fully connected layer. As usual, the motivation for pretraining is that the network is likely to learn on the real task faster if it is pretrained in advance. Note, however, that ResNet-34's pretraining was on images (ImageNet) not sounds.\footnote{We used the pretrained weight file ImageNet1K\_V1.} ResNet-34 was originally designed for 1,000 outputs, that is, classes of images. Since our data set only has four orca behaviors that the network is trying to predict, we modified the ResNet architecture to be four-headed, that is, to have four outputs. 

Capturing fine-temporal-resolution orca behavior data together with sound would be extremely difficult. Wellard~\shortcite{Wellard20:Cold} labeled their sound recordings with all the behaviors observed during the recording period. For this reason, the majority of the sound segments had superfluous labels. This is due to the fact that, while the orcas were doing all of the labeled behaviors during a given recording (which were 51 seconds to 41 minutes long), they may not have been doing all of those behaviors in each sound segment within the recording, so the segments (which were 0.5 to 82.7 seconds long) have superfluous labels. The potentially superfluous labels on the souwaveformss create a difficult classification problem since \textit{there is no ground truth} (except for certain sound segments where the only behavior label was {T}, Table~\ref{ta:frequencies})!

We developed the following approach for dealing with the issue of superfluous behavior labels. We assume that only one behavior was present during each sound segment.\footnote{It is conceivable that this assumption may not be fully accurate for some sound segments in the orca context, but as we will show, we get high classification accuracy with it. Also, it is conceivable that multiple orcas could be producing overlapping or back-to-back vocalizations in a given segment and/or that different orcas in a pod could be exhibiting different behaviors during a segment. However, these are not a problem for our model.} This enables us to leverage recent theory of \textit{partially labeled learning (PLL)}.  In PLL, each training instance may have multiple labels, but only one of them is correct. For PLL, Feng \textit{et al.}~\shortcite{Feng20:Provably} proved that the following loss function is risk consistent. They also introduced a classifier-consistent loss function but showed that especially when using deep learning as the classifier, the risk-consistent loss function performs significantly better in practice. They also showed that the risk-consistent loss function outperforms prior techniques for PLL from the literature~\cite{Feng2019:Partial,Cour11:Learning,Zhang15:Solving,Zhang2017DisambiguationFreePL,Eyke05:Learning}. For these reasons, we use it as the custom loss function for our neural network.  This custom loss function enables the network to learn from training data with superfluous labels. 
\begin{equation}
\hat{R}(f)= \frac{1}{n}\sum_{o=1}^n 
\left(\sum_{i=1}^k\frac{p(y_o=i | x_o)}{\sum_{j\in{Y_o}} p(y_o=j | x_o)} 
{\cal L}(f(x_o),o,i)\right)
\end{equation}
Here the index $o$ is used to sum over instances and the index $i$ to sum over labels. The feature vector (decibel Mel spectrogram in our setting) is $x_o$. The network's prediction for instance $o$ is $y_o$. The label set of training instance $o$ is $Y_o$. The values $p(y=i | x)$ are, of course, not accessible given the data, so we compute them as the softmax'd version of the network's output $f_i(x)$ but only if the label is actually a candidate label in the label set (as Feng \textit{et. al.}~\shortcite{Feng20:Provably} does). 
Formally, the softmax is
\begin{equation}
g_i(x) = \frac{e^{f_i(x)}}{\sum_j e^{f_j(x)}}
\end{equation}
and $p(y=i | x)$ is computed as follows:
\begin{equation}
p(y=i | x) = \begin{cases}
g_i(x) & \mbox{ if } \ i \in Y_o \\
0 & \mbox{ otherwise. } 
\end{cases}
\end{equation}
Finally, ${\cal L}$ is the cross entropy loss of the softmax'd predictions:
\begin{equation}
{\cal L}(f(x_o),o,i) = \begin{cases}
- \log g_i(x_o) & \mbox{ if } \ i \in Y_o \\
0 & \mbox{ otherwise. } 
\end{cases}
\end{equation}

For the optimizer in the neural network's back propagation we used Adam~\cite{Kingma14:Adam}. We changed the learning rate on a schedule. We started with a learning rate of $2 \cdot 10^{-4}$. We decreased the learning rate by a factor of 10 every 10 training epochs.

\section{Experiments}
\label{se:results}

We evaluated our machine learning model over 20 cross-validation repetitions. We split the sound segments (\textit{i.e.}, instances) into a test set and a training set so that 20\% of the data went to the test set and 80\% went to the training set---in a way that the numbers of segments with each of the behavior label combinations were 80-20 proportionate across the training and test set. We also shuffled the segments before assigning them into the test or training set. This means that for each repetition in the cross validation the test and training sets had different instances. We used mini-batches of 10 training instances (sound segments)  for training and testing. We also shuffled the segments within each mini batch. 

\begin{figure}[t]
\begin{center}
    \includegraphics[width = 1.1\columnwidth]{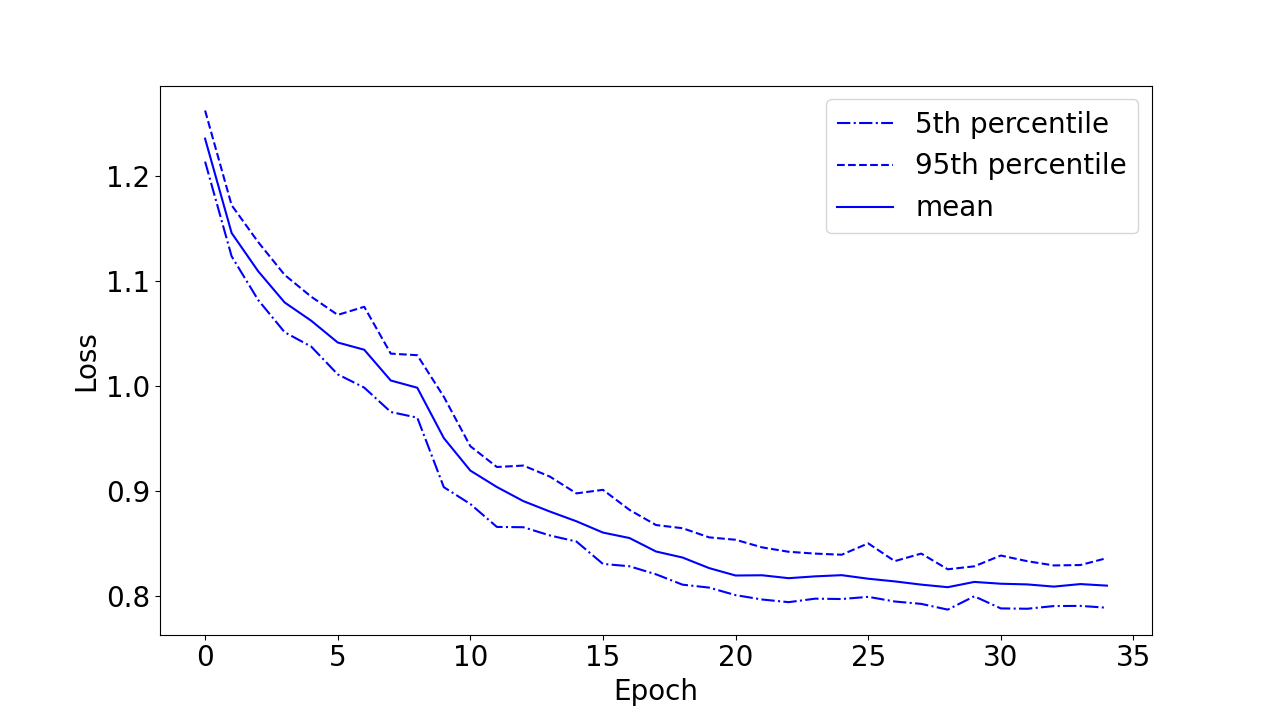}
    \caption{Training set loss $\hat{R}(f)$ as a function of training epochs (with 20-fold cross-validation). Epochs are labeled from 0, so the plot starts after the first training epoch has been completed.}
\label{fi:accgraph}
\end{center}
\end{figure}

\begin{figure}[t]
\begin{center}
    \includegraphics[width = 1.1\columnwidth]{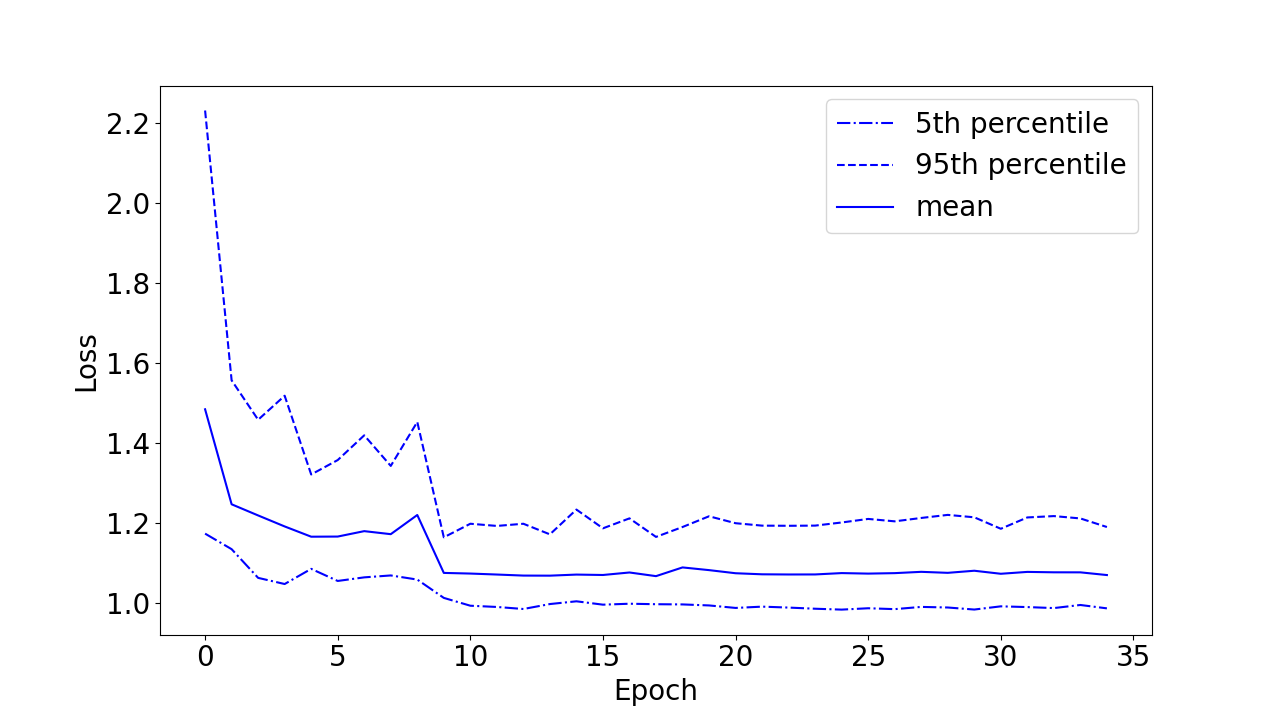}
    \caption{Test set loss $\hat{R}(f)$ as a function of training epochs (with 20-fold cross-validation). Epochs are labeled from 0, so the plot starts after the first training epoch has been completed.}
\label{fi:telossgraph}
\end{center}
\end{figure}

\begin{figure}[t]
\begin{center}
    \includegraphics[width = 1.1\columnwidth]{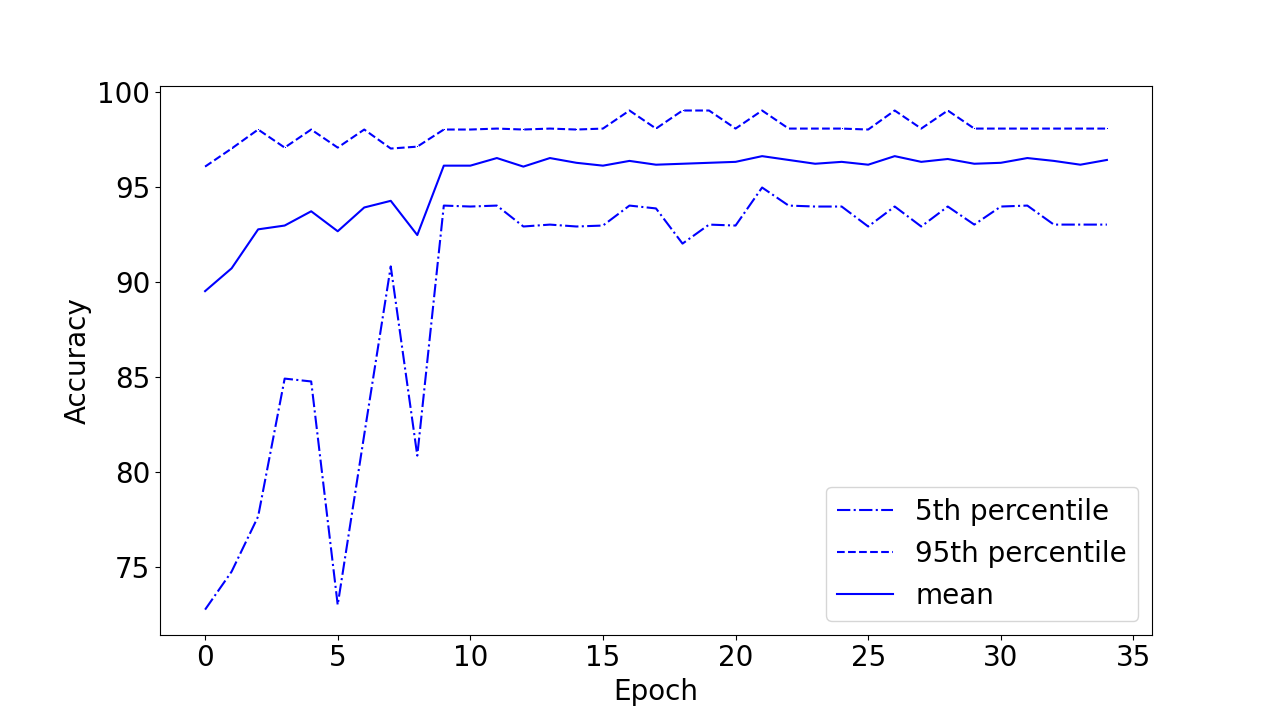}
    \caption{Test set classification accuracy (\%) as a function of training epochs (with 20-fold cross-validation). Epochs are labeled from 0, so the plot starts after the first training epoch has been completed.}
\label{fi:trloss}
\end{center}
\end{figure}

We defined the accuracy on the test set so that if the network's highest-predicted-probability behavior was among the (potentially more than one) labels assigned by Wellard \textit{et al.} for the sound recording from which the sound segment came, the prediction is considered correct. For example, the network could classify a \{T,S,F,M\} file as either \{T\} , \{F\} , \{S\} , or \{M\} , but could only classify a \{T\}  file as \{T\}  in order to get the classification correct. Our machine learning model achieved 96.4\% classification accuracy on the test set, converged to 1.1 loss on the test set, and 0.8 on the training set, shown in Figures~\ref{fi:accgraph}, \ref{fi:telossgraph}, and \ref{fi:trloss}, respectively.

As a performance benchmark and a sanity check, we calculated the accuracy that would be achieved by guessing randomly or by always guessing the same single behavior. We found that guessing uniformly at random would achieve 55.0\% accuracy.  Always guessing T, the most prevalent behavior (with 396 of the segments having T as a possible label), would achieve 76.4\% accuracy, as shown in Table~\ref{ta:guess_accuracies}. 
\begin{table}[!h]
    \centering
    \begin{tabular}{cc}
        \hline
        Guess  & Analytically calculated accuracy \\
        \hline
        Uniform random & 55.0\% \\
        Always guess T & 76.4\% \\
        Always guess F & 74.7\% \\
        Always guess S & 36.1\% \\
        Always guess M & 32.8\% 
    \end{tabular}
    \caption{Sanity check: Accuracy that would result from various behavior guessing schemes based on data from Table~\protect{\ref{ta:frequencies}}.}
    \label{ta:guess_accuracies}
\end{table}

Given that our algorithm achieved 96.4\% accuracy, which is dramatically higher than the 76.4\% achievable by guessing the most prevalent class, we can see that the orca sound segments predict orca behavior. Even at bottom 5\% percentile, our model achieved 93.0\% accuracy, seen in Figure ~\ref{fi:accgraph}, which is higher than the 76.4\% accuracy that the model would achieve if guessing the most prevalent class. Therefore, our model is better than always guessing the most prevalent class with more than 95\% statistical significance. These results strongly show that orca sound segments contain indications of behavior. The fact that orcas are communicating behavior indicates that they use a semantic language. 

\section{Examples of What the System Learned}
\label{se:examples_of_what_the_system_learned}
\begin{figure}[h!]
\begin{center}
    \includegraphics[width = \columnwidth]{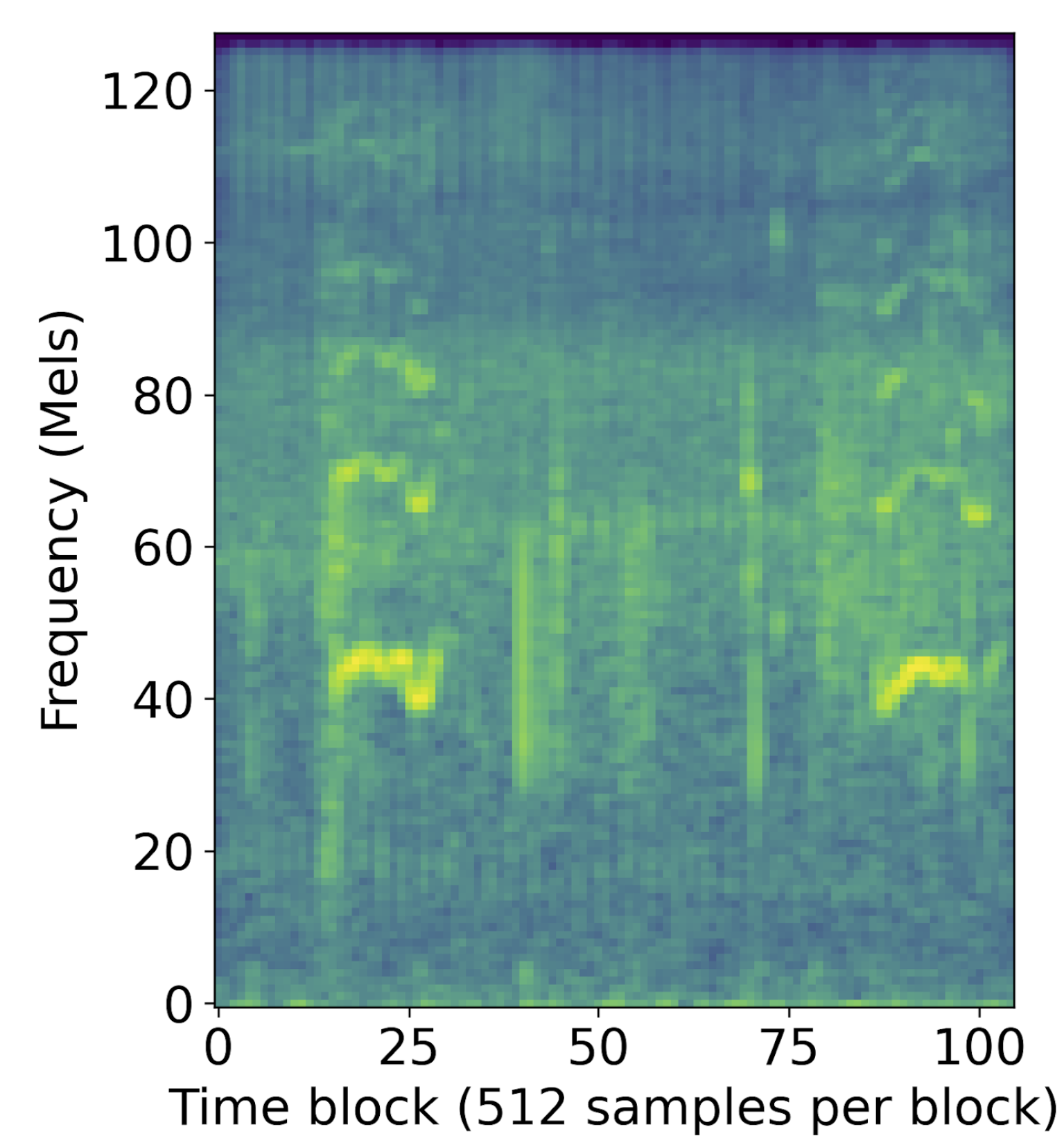}
    \caption{Resampled decibel Mel spectrogram of a 2-second-long \{T,F,S,M\} segment which our system classifies.}
    \label{fi:exampleC}
\end{center}
\end{figure}

\begin{figure*}[t]
\begin{center}
    \includegraphics[width = \textwidth]{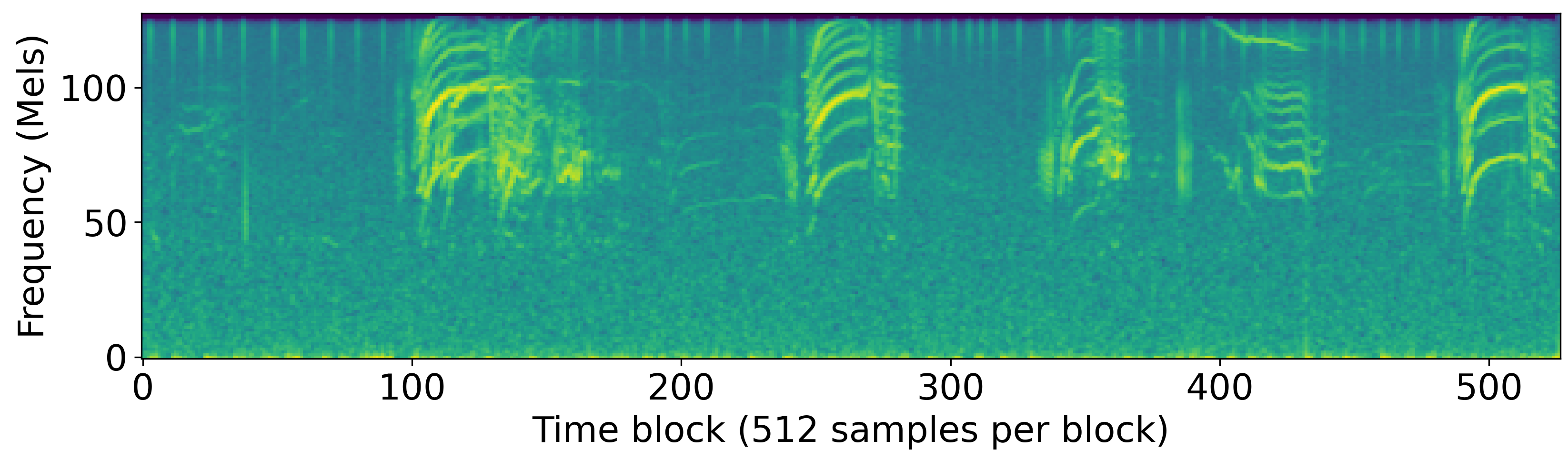}
    \caption{Resampled decibel Mel spectrogram of a 12.2-second-long \{T,F\} segment which our system classifies.}
    \label{fi:exampleC1}
\end{center}
\end{figure*}

In this section, we show two examples of sound segments that our system classifies. We show resampled decibel Mel spectrograms of these segments in Figures~\ref{fi:exampleC} and~\ref{fi:exampleC1}, respectively. We show the softmax'd predictions that our system outputs in Tables ~\ref{ta:exampleClass} and ~\ref{ta:exampleClass1}), respectively.

\begin{table}[!h]
    \centering
    \begin{tabular}{cc}
        \hline
        Class  &  Prediction by the system\\
        \hline
        T & 49.5\% \\
        F & 34.4\% \\
        S & 5.9\% \\
        M & 10.1\% \\
    \end{tabular}
    \caption{Our system's softmax prediction values for the \{T,F,S,M\} sound in Figure~\protect{~\ref{fi:exampleC}}.}
    \label{ta:exampleClass}
\end{table}
Figure~\ref{fi:exampleC} is a resampled decibel Mel spectrogram of a segment which is spuriously labeled as \{T,F,S,M\} in the input data, that is, the input data has no information about the correct behavior: having all the labels in the label set means that the input data has no information on that instance. 
Our system classifies this sound as indicating behavior T. This is shown in Table~\ref{ta:exampleClass}. Our system assigns probability 49.5\% on the behavior being T out of all four behaviors. Although the prediction may seem relatively low, the network was very sure that the behavior was not S, showing that the system was able to find predictive structures within the segment. It is also possible that certain orca behaviors have structures within the sound segments that are more similar to certain behaviors than to others. In this example, the network was 49.5\% sure that the behavior was T and 34.4\% sure that the behavior was F. This may suggest that the foraging and traveling behaviors have more similarities in their vocalization structures than traveling and socializing. 

Figure~\ref{fi:exampleC1} is a resampled decibel Mel spectrogram of a sound segment which had label set \{T,F\} in the input data. 
The system assigns probability 75.8\% on the behavior being F, shown in Table ~\ref{fi:exampleC1}. The system predicted that there was a 8.7\% probability that the behavior was T, which shows that our system is able to separate traveling from foraging with great certainty in this case even though their associated sounds might have similar aspects as suggested by the first example in this section.
\begin{table}[h!]
    \centering
    \begin{tabular}{cc}
        \hline
        Class  &  Prediction by the system\\
        \hline
        T & 8.7\% \\
        F & 75.8\% \\
        S & 13.2\% \\
        M & 2.2\% \\
    \end{tabular}
    \caption{Our system's softmax prediction values for the \{T,F\} sound in Figure~\protect{~\ref{fi:exampleC1}}.}
    \label{ta:exampleClass1}
\end{table}

\section{Conclusions and Future Research}
\label{se:conclusions}

\textit{Orcinus orca} (killer whales) have complex vocalizations that use multiple frequencies simultaneously, vary the frequencies, and vary their volumes. To our knowledge, it was not known whether calls or longer sound \textit{segments}---call \textit{sequences} or even more general structures---constitute a \textit{semantic} language. That is, do orcas communicate meaning? 

This lack of knowledge stems from orcas living mostly under water and moving quickly, so sound recordings are rarely accompanied by other data that would support such reasoning. We used a recent orca sound recording collection that is rare in the sense that it has such auxiliary data~\cite{Wellard20:Dataset,Wellard20:Cold}. In particular, it has partially labeled behavior data. By carefully combining and tailoring select modern machine learning techniques, we showed that the sound segments can be used to predict orca behavior with 96.4\% accuracy. This revealed the highly  predictive properties that orca sound segments have when it comes to classifying behavior. The fact that the sound segments can be used to classify behavior suggests that orcas convey meaning through their vocalizations, that is, they use a semantic language.

Our most significant task in classifying orca behaviors based on sound segments was developing a machine learning system that learns with partially labeled orca sound segment data. When recording the orcas, Wellard \textit{et al.} recorded all of the orca behaviors observed. We segmented the recordings so that the segments were isolated. In a given segment, an orca may not be exhibiting all of the behaviors that were observed during the recording from which the segment was taken. Therefore, we were left with segments that have potentially superfluous labels. Our learning system consists of a tailored sound preprocessing pipeline, a slightly modified ResNet-34 convolutional neural network, and a custom loss function for training it with partially labeled data.

To our knowledge, this paper is the first to use partially labeled learning to study animal vocalizations, first to use machine learning to analyze orca sound segments beyond individual calls,  first to predict orca 
behavior from their vocalizations, and first to successfully explore the link between orca vocalizations and behavior.

This work and system could help marine biologists study orca behavior with greater capacity. Currently, observing orca behavior is quite difficult since orcas live under water and their behaviors may not be obvious to an above-water observer. Our system would be helpful for marine biologists because it would allow marine biologists to record orca vocalizations and identify orca behaviors with greater ease and efficiency. Given the growing threats of global warming on marine ecosystems, our system  could greatly help marine biologists and conservationists who are trying to solve the climate-related problems that orcas face. For example, our system would make it possible to find out whether orcas are foraging for food at a decreasing rate over a period of time. It might also in the long run help explain why orca pods have recently started attacking recreational vessels. Our system could also have safety applications and other practical applications.

Our system could also aid further research into whale language. For example, one could use the segments which our algorithm classified as a certain behavior and compare the structure of segments where the orcas are exhibiting different behaviors to study the differentiating features. This would give researchers an insight into the potential grammar structure of orca language. Such research may also help with the task of translating orca language. For example, if a grammar structure is discovered using our system, one could use the behavior data and the structure data to get a better picture of what orcas are saying. Our algorithm may also contribute to the development of more data sets with behavior labels which would allow researchers working on orca language to have more data to work with. This could allow for greater ease and creativity when studying orca language with machine learning.



%

\bibliographystyle{named}
\bibliography{orca}
\end{document}